\documentclass[aps,prb,preprint,superscriptaddress,floatfix]{revtex4}

\usepackage{graphicx,color}
\graphicspath{{figs/}}
\begin{document}
\title{Orientation Dependent Thermal Conductance in Single-Layer MoS$_{2}$}
\author{Jin-Wu Jiang  \footnote{Correspondence to: jinwu.jiang@uni-weimar.de (JWJ)}}
    \affiliation{Institute of Structural Mechanics, Bauhaus-University Weimar, Marienstr. 15, D-99423 Weimar, Germany}
\author{Xiaoying Zhuang}
    \affiliation{Department of Geotechnical Engineering, College of Civil Engineering, Tongji University, 1239  Siping Road, Shanghai 200092, China}
\author{Timon Rabczuk \footnote{Correspondence to: timon.rabczuk@uni-weimar.de (TR)}}
    \affiliation{Institute of Structural Mechanics, Bauhaus-University Weimar, Marienstr. 15, D-99423 Weimar, Germany}
    \affiliation{School of Civil, Environmental and Architectural Engineering, Korea University, Seoul, South Korea }
%\date{22 December 2009}
\date{\today}
\begin{abstract}
We investigate the thermal conductivity in the armchair and zigzag MoS$_{2}$ nanoribbons, by combining the non-equilibrium Green's function approach and the first-principles method. A strong orientation dependence is observed in the thermal conductivity. Particularly, the thermal conductivity for the armchair MoS$_{2}$ nanoribbon is about 673.6 Wm$^{-1}$K$^{-1}$ in the armchair nanoribbon, and 841.1 Wm$^{-1}$K$^{-1}$ in the zigzag nanoribbon at room temperature. By calculating the Caroli transmission, we disclose the underlying mechanism for this strong orientation dependence to be the fewer phonon transport channels in the armchair MoS$_{2}$ nanoribbon in the frequency range of [150, 200]~{cm$^{-1}$}. Through the scaling of the phonon dispersion, we further illustrate that the thermal conductivity calculated for the MoS$_{2}$ nanoribbon is esentially in consistent with the superior thermal conductivity found for graphene.
\end{abstract}

%\pacs{65.80.Ck, 63.22.Np, 68.65.-k}
%\keywords{thermal conductivity, single-layer MoS$_{2}$, orientation dependence, anisotropy}
\maketitle
\pagebreak

The two-dimensional single-layer MoS$_{2}$ (SLMS) has drawn considerable interest in recent years, owning to its direct band gap and graphene-like layered honeycomb lattice structure.\cite{JoensenP,HelvegS,MakKF,LeeC,RadisavljevicB2011nn,PopovI,AtacaC,SanchezAM,SahooS,YinZ,ChangK,AtcaC,RadisavljevicB2012apl,Castellanos-GomezA,ColemanJN,LiuKK,LeD,BrivioJ} The band gap in the SLMS is a great advantage over graphene, which has zero band gap without the help of strain or other gap-opening techniques. Thus, the SLMS becomes a promising alternative to graphene in many electronic fields. For this purpose, many efforts have been devoted to comparatively investigate various properties of SLMS and graphene. It was found that the interface between the SLMS and the gold metal is inefficient for electron injection; while the Ti metal can greatly improve the efficiency.\cite{PopovI} The mechanical and electronic properties of the SLMS have been investigated by first-principles calculations, where the defects were found to play an important role.\cite{AtacaC} S\'{a}nchez and Wirtz studied the phonon properties of the SLMS, using the first-principles calculations.\cite{SanchezAM}

In a very recent experiment, Sahoo {\it et.al} measured the thermal conductivity in the SLMS through analyzing the laser power dependence of the Raman mode.\cite{SahooS} The obtained thermal conductivity is 52 Wm$^{-1}$K$^{-1}$ at room temperature. The thermal conductivity is an important parameter for possible applications of the SLMS. High thermal conductivity is helpful to deliver heat away from the electronic device, which is good to prevent heating induced breakdown of the device. For thermoelectric applications, lower thermal conductivity can yield in higher figure of merit ZT.\cite{GoldsmidHJ1964,GoldsmidHJ1986,NolasGS} Hence, it is useful to provide an accurate prediction for the thermal conductance value, which can be used to predict the upper limit of the thermal conductivity in a pure SLMS without any phonon decaying mechanism. The predicted thermal conductivity value can serve as an upper limit for the values from the experimental measurement or other theoretical calculations.

In this paper, we comparatively study the thermal conductivity in both armchair and zigzag MoS$_{2}$ nanoribbons (MSNRs), by using the ballistic non-equilibrium Green's function (NEGF) formalism and the first-principles calculations. We find that the thermal conductivity is strongly dependent on the orientation. The room temperature thermal conductivity in the armchair MSNR is about 25\% lower than that of the zigzag MSNR. The Caroli transmission function shows that the origin for the strong orientation dependence is because of the much less phonon transport channels in the armchair MSNR in the frequency range of [150, 200]~{cm$^{-1}$}.

\section*{Results}
In the {\it ab initio} calculation, we use the SIESTA package\cite{SolerJM} to optimize the structure of the SLMS. The local density approximation is applied to account for the exchange-correlation function with Ceperley-Alder parametrization\cite{CeperleyDM} and the double-$\zeta$ basis set orbital is adopted. During the conjugate-gradient optimization, the maximum force on each atom is smaller than 0.01 eV\AA$^{-1}$. A mesh cut off of 120 Ry is used. Periodic boundary condition is applied in the heat current direction and the in-plane transverse direction. The free boundary condition is applied to the out-of-plane direction by introducing a vacuum space of 26~{\AA}. There are two types of calculations in this work. In the calculation of the phonon spectrum, a small unit cell is used and an $8\times 1\times 1$ Monkhorst-Pack k-grid is chosen for the sampling of the quasi-one-dimensional Brillouin zone. In the investigation of the thermal transport, a large unit cell is calculated and the Gamma point $k$ sampling is adopted.

The relaxed structures for armchair and zigzag MSNRs are shown in Fig.~\ref{fig_cfg}. The Mo-S bond length is 2.414~{\AA}, which is between the value 2.382~{\AA} from S\'{a}nchez {\it et.al}\cite{SanchezAM} and 2.42~{\AA} from Ataca {\it et.al}.\cite{AtacaC} There are 36 (18) atoms in the smallest translation unit cell in the armchair (zigzag) MSNR. Fig.~\ref{fig_phonon} compares the phonon dispersion in the armchair (left) and zigzag (middle) MSNRs. Two smallest translation unit cells are included for the armchair MSNR in the calculation of the phonon dispersion. For the zigzag MSNR, four smallest translation unit cells have been used. Hence there are 72 atoms involved in the calculation, resulting in 216 phonon branches in both panels. The wave vector $k$ is along the heat current direction and in the unit of $2\pi/a$, with $a=10.705$~{\AA} or 12.364~{\AA} in the armchair or zigzag MSNR, respectively. An obvious band gap around 300~{cm$^{-1}$} is found in the phonon spectra for both armchair and zigzag MSNRs. Similar band gap has been found in some existing works.\cite{AtacaC,SanchezAM} The three acoustic phonon branches from the armchair and zigzag MSNRs are compared in the right panel. We note that the $x$ axis here is in the unit of \AA$^{-1}$, which is different from the left and middle panels in the same figure. The three acoustic phonon branches in the armchair MSNR are considerably higher than that in the zigzag MSNR in the low-frequency range. In other words, the sound velocities in the armchair MSNR are higher than that in the zigzag MSNR. This can be simply understood from the quite different configuration of these two orientations as shown in Fig.~\ref{fig_cfg}. The top panel in Fig.~\ref{fig_cfg} shows that the projections of two Mo-S bonds are along the phonon transport direction (horizontal), which can enhance the mechanical properties of the system in the horizontal direction. However, the bottom panel in Fig.~\ref{fig_cfg} shows the projections of two Mo-S bonds are perpendicular to the phonon transport direction, so these two bonds have few contribution for the mechanical strength in the horizontal direction. As a result, the acoustic sound velocities in the armchair MSNR are higher than that in the zigzag MSNR. It should be noted that the out-of-plane transverse branch should behave flexibly due to the two-dimensional nature of the SLMS. This flexure nature disappears in present calculation, due to the loss of the rigid rotational invariance symmetry in the {\it ab initio} calculation.\cite{JiangJW2006}

There are various options to calculate the thermal transport. Classical results can be obtained from the molecular dynamics simulation of the thermal transport, where the phonon-phonon scattering dominates the transport process. From the theoretical point of view, it is a big challenge to provide accurate prediction for the thermal conductivity, because the samples in the experiment always possess various unpreventable defects. Hence, a more practical task is to provide an accurate (quantum) prediction for the upper limit of the thermal conductivity. For this purpose, we would rather apply the ballistic NEGF method.\cite{WangJSnegf,JiangJW2011negf} It is based on quantum mechanics. The phonon-phonon scattering is ignored in the ballistic transport region, which is actually quite reasonable for low-dimensional nano-materials. For instance, in graphene, it has been found that the low-frequency phonons can have very large mean-free path in graphene.\cite{NikaDL2009prb,NikaDL2012jpcm} The combination of the NEGF and the {\it ab initio} calculation can provide us an accurate upper limit for the thermal conductivity.\cite{JiangJW2011defect} Please see the section Method for more details.

\section*{Discussion}
Fig.~\ref{fig_kappa} top panel shows the transmission functions for the armchair and zigzag MSNRs. These functions exhibit some regular steps, due to the absence of phonon-phonon scattering. In the ballistic transport, $\sigma$ is proportional to the cross-sectional area, since there are more channels available for heat delivery in thicker nanoribbons. However, the thermal conductance from equation~(\ref{eq_conductance}) does not depend on the length of the system. $\sigma$ can be used to get the thermal conductivity ($\kappa$) of a MSNR with arbitrary length $L$: $\kappa=\sigma L/s$, where $s$ is the cross-sectional area. We have assumed the thickness of the MSNR to be 6.033~{\AA}, which is the space between two adjacent layers in the bulk MoS$_{2}$.\cite{SanchezAM} This thickness value is the same for both armchair and zigzag MSNRs, so its value does not affect our comparison for their thermal conductivity. It is quite obvious that the thermal conductivity does not depend on the cross section. It means that the thermal conductivity in the MSNR does not depend on its width, since the thickness is a constant. On the other hand, the thermal conductivity in ballistic region increases linearly with increasing length, which has been observed in the thermal conductivity of two-dimensional graphene.\cite{BalandinAA2008,NikaDL2009prb,NikaDL2009apl,Balandin2011nm,JiangJW2009direction} In the experiment, the thermal conductivity is measured for a MSNR sample of dimension around $\mu$m.\cite{SahooS} Hence, we will predict the thermal conductivity for a MSNR of length $L=1.0$~{$\mu$m}.

In recent years, a huge amount of studies have shown that there should be an additional regime between the diffusive and ballistic transport regimes.\cite{YangN2010nt} As stimulated by the super diffusion transport regime, we feel that it is useful to illustrate the size dependence for the thermal conductivity besides the thermal conductance for the nano-structure, MoS$_{2}$ nanoribbon, although our study is in the ballistic regime. Furthermore, our prediction for the size dependence of the thermal conductivity in the MoS$_{2}$ nanoribbon is of practical significance, considering the intense interests currently in the size effects on the thermal transport in nano-materials. Our ballistic prediction shows that the thermal conductivity increases linearly with increasing length. This phenomenon is well-known, as it is in the ballistic regime. However, the exact value for the thermal conductivity will be the upper limit for the system with a particular length. This is good for future studies in this field.

Fig.~\ref{fig_kappa} bottom panel gives the thermal conductivity for both armchair and zigzag MSNRs of 1.0~{$\mu$m} in length. A strong orientation dependence is revealed in the thermal conductivity. The thermal conductivity in the armchair MSNR is smaller than that of the zigzag MSNR in the whole temperature range. The orientation-induced anisotropy is about 25\% at room temperature. This anisotropy is much stronger than that in the graphene. In the ballistic region, the orientation-induced anisotropy is only about 1\% in the single-layer graphene.\cite{JiangJW2009direction}

The top panel in the figure discloses the origin for this strong anisotropy. It is because the transmission function in the armchair MSNR is much smaller than that of the zigzag MSNR in the frequency range [150, 200]~{cm$^{-1}$}. It means that the phonon transport channels are fewer in this frequency domain for armchair MSNR, leading to much smaller thermal conductivity in the armchair MSNR, although the three acoustic velocities in the armchair MSNR are higher as shown in Fig.~\ref{fig_phonon} right panel. Particularly, the room-temperature thermal conductivity is 673.6 and 841.1 Wm$^{-1}$K$^{-1}$ in armchair and zigzag MSNRs. These values are about one third of the superior thermal conductivity value (around 2000 Wm$^{-1}$K$^{-1}$) in the graphene,\cite{BalandinAA2008} because the overall phonon spectrum in the MSNR ([0, 500]~cm$^{-1}$) is scaled by a factor of one third from the phonon spectrum in the graphene ([0, 1600]~cm$^{-1}$). The experiment value is 52 Wm$^{-1}$K$^{-1}$, which is far bellow our quantum upper limit. It indicates strong phonon scattering or substrate-induced phonon leaking effects in the experimental samples, as suggested by the authors in Ref.~\onlinecite{SahooS}.

We should stress that our theoretical value serves as an upper limit for the thermal conductivity of the 1.0~{$\mu$m} MSNRs. If the experimental samples are of high quality, then the measured thermal conductivity should approach 673.6 or 841.1 Wm$^{-1}$K$^{-1}$ from lower side.

%\section{conclusion}
In conclusion, we have applied the ballistic NEGF approach to predict the upper thermal conductivity value in the armchair and zigzag MSNRs. The force constant matrix is calculated from the first-principles method. A strong orientation dependence is found in the thermal conductivity. More specifically, the thermal conductivity in the armchair MSNR is about 25\% smaller than the zigzag MSNR at room temperature. The Caroli transmission function discloses the origin for this strong anisotropy to be the much less transport channels in the armchair MSNR in the frequency range of [150, 200]~{cm$^{-1}$}.

\section*{Methods}
In this section, we outline some key steps in applying the NEGF approach for the thermal transport. More details about the NEGF approach can be found elsewhere.\cite{WangJSnegf,JiangJW2011negf} In the NEGF approach, the thermal conductance is calculated by the Landauer formula:
\begin{eqnarray}
\sigma & = & \frac{1}{2\pi}\int d\omega\hbar\omega T[\omega]\left[\frac{\partial n(\omega,T)}{\partial T}\right],
\label{eq_conductance}
\end{eqnarray}
where $\hbar$ is the Planck's constant. $n(\omega,T)$ is the Bose-Einstein distribution function. The transmission $T[\omega]$ is obtained from the Caroli formula:
\begin{eqnarray}
T[\omega] & = & {\rm Tr}\left(G^{r}\Gamma_{L}G^{a}\Gamma_{R}\right),
\end{eqnarray}
where $G^{r}$ is the retarded Green's function. $G^{a}=\left(G^{r}\right)^{\dagger}$ is the advanced Green's function and $\Gamma_{L/R}$ is the self-energy.

The retarded self-energy of the leads are obtained by:
\begin{eqnarray}
\Sigma_{\alpha}^{r} & = & V^{C\alpha}g_{\alpha}^{r}V^{\alpha C},
\end{eqnarray}
which carries the coupling information between leads and center region. The surface Green's function, $g_{\alpha}^{r}$, is calculated using the iterative approach.\cite{SanchoMPL} Then calculate the $\Gamma$ function:
\begin{eqnarray}
\Gamma_{\alpha} & = & i\left(\Sigma_{\alpha}^{r}-\Sigma_{\alpha}^{a}\right)=-2Im\Sigma_{\alpha}^{r}.
\end{eqnarray}

The retarded Green's function for the center region connected with leads is then obatained:
\begin{eqnarray}
G^{r} & = & \left[\left(\omega+i\eta\right)^{2}I-K^{C}-\Sigma_{L}^{r}-\Sigma_{R}^{r}\right]^{-1}.
\end{eqnarray}

The phonon dispersions are obtained by solving the eigenvalue problem of the dynamical matrix, which is derived from the first-principles calculation. The dynamical matrix is obtained by $K_{ij}=\partial^{2}V/\partial x_{i}\partial x_{j}$, where $V$ is the potential from Siesta and $x_{i}$ is the position of the i-th degree of freedom. This formula is realized numerically by calculating the energy change after a small displacement of the i-th and j-th degrees of freedom.

\textbf{Acknowledgements} The work is supported by the Grant Research Foundation (DFG). X. Y. Z acknowledges the support by the National Basic Research Program of China (973 Program: 2011CB013800) and the Program for Changjiang Scholars and Innovative Research Team in University (PCSIRT, IRT1029).

\textbf{Author contributions} J.W.J designed the project, performed the calculations and wrote the paper. X.Y.Z discussed the results. R.T discussed the results. All authors commented on the manuscript.

\textbf{Competing financial interests} The authors declare no competing financial interests.

%\bibliographystyle{nature}
%\bibliography{biball}

\pagebreak

\begin{figure}[htpb]
%  \begin{center}
%    \scalebox{1.0}[1.0]{\includegraphics[width=8cm]{cfg.eps}}
%  \end{center}
  \caption{The relaxed structure of the armchair (top) and zigzag (bottom) MSNRs. Top and side views are shown for each system. Atoms are denoted by blue (Mo) and red (S). The translation unit cells are highlighted on the left end.}
  \label{fig_cfg}
\end{figure}

\begin{figure}[htpb]
%  \begin{center}
%    \scalebox{1.5}[1.5]{\includegraphics[width=8cm]{phonon.eps}}
%  \end{center}
  \caption{Phonon dispersion in armchair (left) and zigzag (middle) MSNRs. Insets show the unit cells used in the calculation of the phonon spectrum. The three acoustic phonon branches in armchair and zigzag systems are compared in the right panel. We note the band gap around 200~{cm$^{-1}$} in the spectrum.}
  \label{fig_phonon}
\end{figure}

\begin{figure}[htpb]
%  \begin{center}
%    \scalebox{1.0}[1.0]{\includegraphics[width=8cm]{kappa.eps}}
%  \end{center}
  \caption{Top: the transmission functions for armchair and zigzag MSNRs. Note obvious less transmission channels in the armchair MSNR in [150, 200]~{cm$^{-1}$}. Bottom: the thermal conductivity ($\kappa$) obtained from thermal conductance ($\sigma$) through their relationship, $\kappa=\sigma L/s$, with cross-sectional area $s=wh$. The thickness $h=6.033$~{\AA}, and the length $L=1$~$\mu$m. $\sigma$ is calculated by the Landauer formula.}
  \label{fig_kappa}
\end{figure}

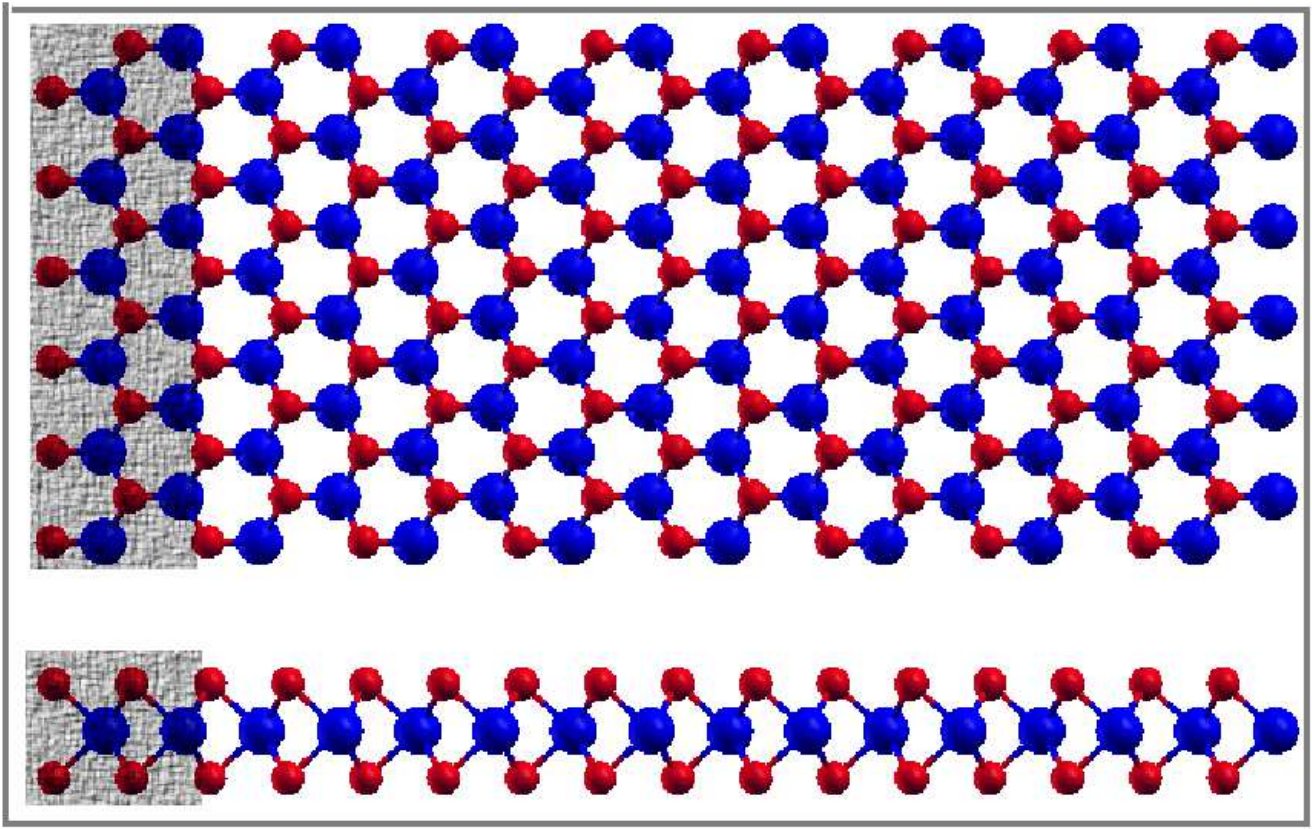
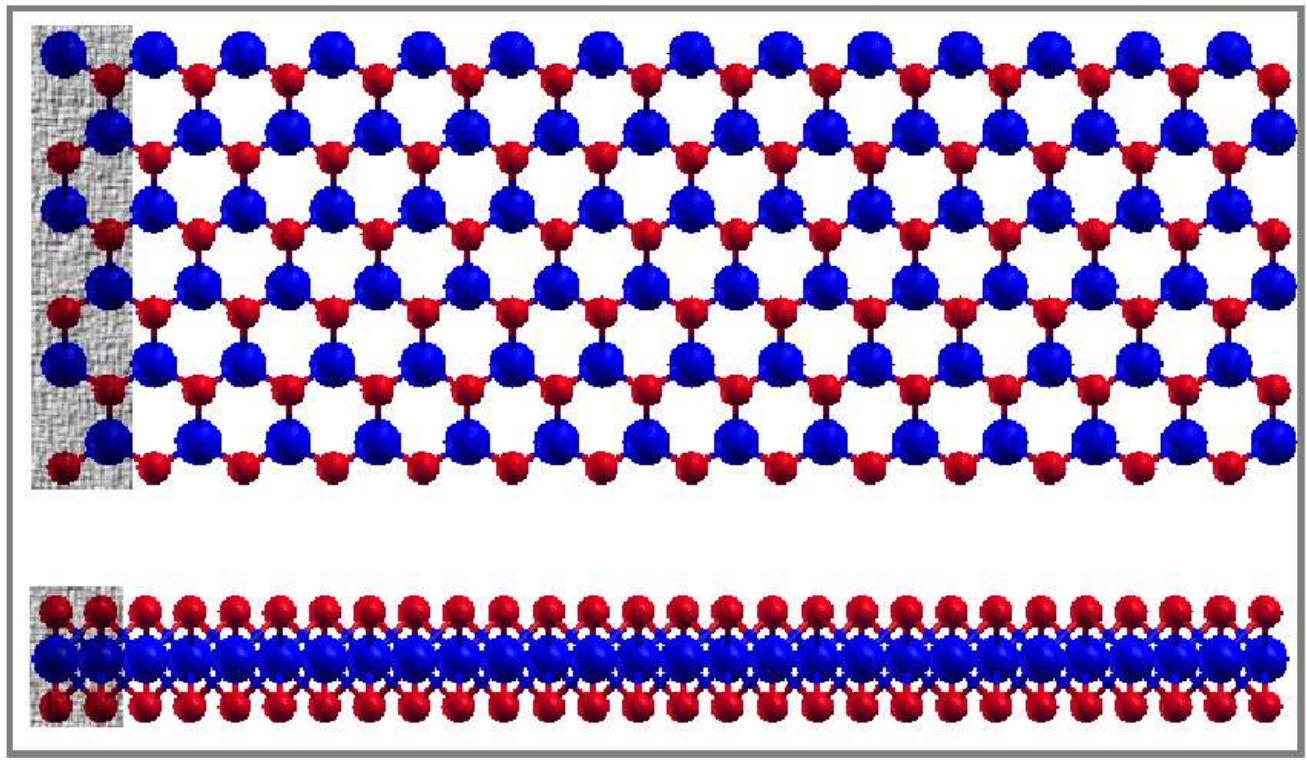
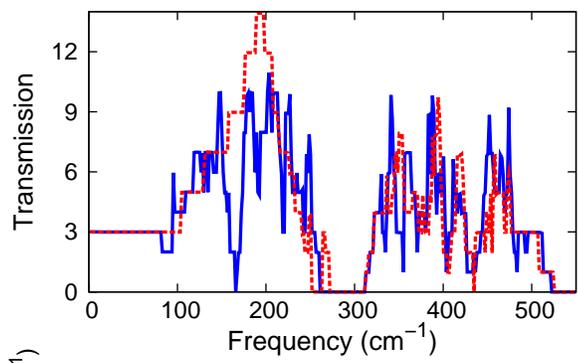
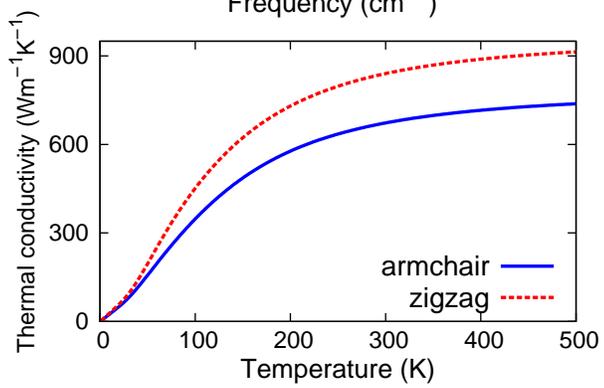
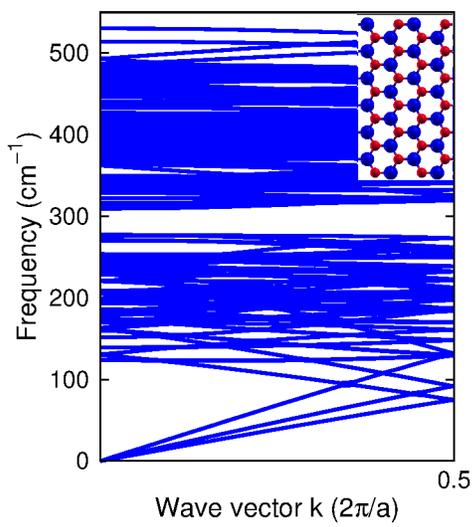
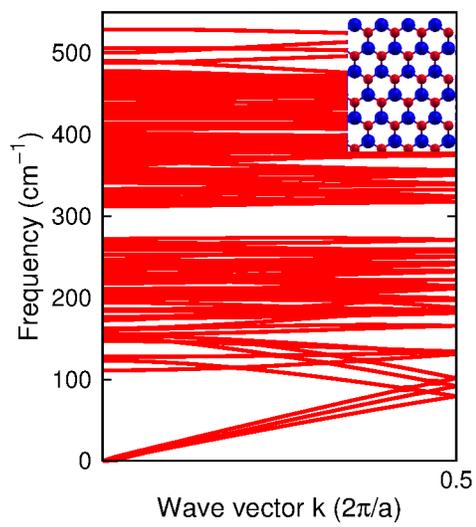
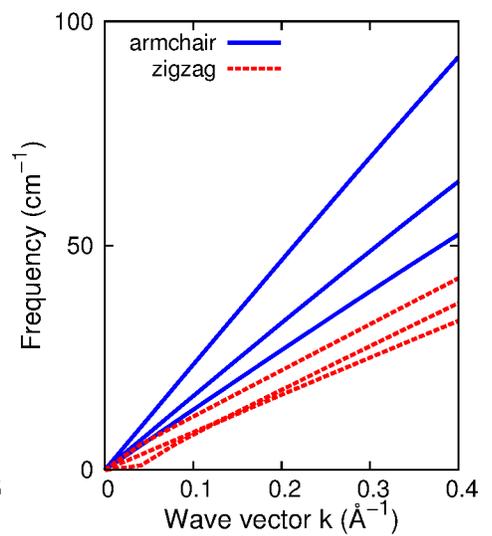
\end{document}